\documentclass[prb,aps,twocolumn,superscriptaddress]{revtex4}

\usepackage{graphicx}% Include figure files
\usepackage{dcolumn}% Align table columns on decimal point
\usepackage{bm}% bold math

\begin{document}

\title{Aspect-ratio dependence of the spin stiffness
of a two-dimensional XY model}

\author{R.~G.~Melko}
%\email{rgmelko@physics.ucsb.edu}
\affiliation{Department of Physics, University of California Santa
Barbara,
California 93106}

\author{A.~W.~Sandvik}
%\email{asandvik@abo.fi}
\affiliation{Department of Physics, University of California Santa
Barbara,
California 93106}
\affiliation{Department of Physics, {\AA}bo Akademi University,
Porthansgatan 3, FIN-20500 Turku, Finland}

\author{D.~J.~Scalapino}
%\email{djs@vulcan2.physics.ucsb.edu}
\affiliation{Department of Physics, University of California Santa
Barbara,
California 93106}

\date{\today}

\begin{abstract}

We calculate the superfluid stiffness of 2D lattice hard-core bosons
at half-filling (equivalent to the $S=1/2$ XY-model)
using the squared winding number quantum Monte Carlo estimator. For $L_x
\times L_y$ lattices with aspect ratio $L_x/L_y=R$, and $L_x,L_y \to \infty$,
we confirm the recent prediction [N. Prokof'ev and B.V. Svistunov, Phys.
Rev. B {\bf 61}, 11282 (2000)] that the finite-temperature
stiffness parameters $\rho^W_x$ and $\rho^W_y$ 
determined from the winding number differ from each other
and from the true superfluid density $\rho_s$. 
Formally, $\rho^W_y \rightarrow \rho_s$ in the limit in which $L_x
\rightarrow \infty$ first and then $L_y \rightarrow \infty$.
In practice we find that $\rho^W_y$ converges exponentially
to $\rho_s$ for $R>1$.
We also confirm that for 3D systems, $\rho^W_x = \rho^W_y = \rho^W_z =
\rho_s$ for any $R$.
In addition, we determine the Kosterlitz-Thouless transition temperature 
to be $T_{KT}/J=0.34303(8)$ for the 2D model.

\end{abstract}

\maketitle

\section{Introduction}

In the usual Kosterlitz-Thouless \cite{KTxx} description of a 2D superfluid,
\cite{BishopReppy}
excitations of bound vortex-antivortex pairs renormalize the superfluid
density $\rho_s(T)$. 
Here, $\rho_s(T)$ is the coefficient of the square of the gradient of the 
phase in the free energy expression that governs long-wavelength
phase fluctuations.
Above a critical temperature $T_{KT}$, the vortex pairs unbind
and $\rho_s(T)$ drops to zero. The well-known Nelson-Kosterlitz formula,
\cite{NKxx}
\begin{equation}
T_{KT} = \frac{\pi}{2}\ \rho_s\, (T_{KT}),
\label{tkt}
\end{equation}
relates $T_{KT}$ to the discontinuity in the superfluid density
$\rho_s(T_{KT})$ at $T_{KT}$. Using quantum Monte Carlo simulations in a
real-space basis, one can calculate a stiffness parameter 
$\rho^W_{\alpha}(T)$, related to $\rho_s(T)$, using a squared winding
number estimator. \cite{stiffness} On an $L_x\times L_y$ torus, we can define 
two stiffness parameters (or helicity moduli)
\begin{eqnarray}
\rho^W_x(T) & = & T \frac{L_x}{L_y}\langle W^2_x\rangle , \\
\label{two}
\rho^W_y(T) & = & T \frac{L_y}{L_x}\langle W^2_y\rangle ,
\label{three}
\end{eqnarray}
where the integer winding numbers $W_\alpha$, $\alpha=x,y$, are defined
according to
\begin{equation}
W_\alpha = \frac{1}{L_\alpha}\int\limits_0^\beta d\tau J_\alpha (\tau),
\end{equation}
where $J_\alpha$ is the boson current operator and $\beta=1/T$.  Typically,
systems with aspect ratio $R=L_x/L_y=1$ have been studied,
and in the limit where $L_x$ and $L_y$ become infinite, one might
expect that $\rho^W_{\alpha}(T)$ is equal to $\rho_s(T)$.
However, recently Prokof'ev and Svistunov \cite{PSxx} have noted that 
in addition to the Kosterlitz-Thouless vortex-antivortex pairs, 
topological excitations present for an $L_x \times L_y$ torus
lead to a renormalization of the stiffness parameters.  In this case,
at finite temperature, even in the limit $L_x,L_y \to \infty$, 
$\rho^W_x (T)$ and $\rho^W_y(T)$ depend upon the aspect ratio and in 
general do not equal the superfluid density $\rho_s(T)$. \cite{DJSxx} 
In particular, for $L_x=L_y$, $\rho^W_x(T_{KT})/\rho_s(T_{KT}) \approx
0.9998$. Hence, calculations of $T_{KT}$ utilizing Eq.~(\ref{tkt}) and assuming
$\rho^W_x = \rho_s$ can be expected to be affected by a small 
(typically negligible) systematic error.  

In this paper we present a quantum Monte Carlo study of the aspect ratio
dependence of $\rho^W_x$ and $\rho^W_y$, organized as follows.  In
Section \ref{PSsec} we review the argument of Ref.~\onlinecite{PSxx}
and summarize their conclusions.  Section \ref{MCsec} contains the
results of our quantum Monte Carlo study on the $S=1/2$ XY-model.  In
\ref{23Dsec} we show the dependence of $\rho^W_x$ and
$\rho^W_y$ on $R$ for 2D and 3D square lattices, and illustrate the
need for an accurate, independent estimate of $T_{KT}$ for this model
to use as a benchmark.
In Section \ref{WMsubsec} we use the Weber-Minnhagen RG scaling
relation \cite{WMxx} on systems with $R=1$ and 4 to obtain a very accurate 
estimate of $T_{KT}$. In Section \ref{rhoRATsec} we explore a
method of calculating $T_{KT}$ using the predicted aspect ratio dependence of 
$\rho^W_x$/$\rho^W_y$, and illustrate the need for precise finite-size scaling 
of the data in order to obtain satisfactory agreement with the benchmark
$T_{KT}$.  Once the finite-size scaling behavior of
$\rho^W_x$/$\rho^W_y$ is understood, it is straightforward to confirm the 
conclusions of Ref.~\onlinecite{PSxx} using our data.

\begin{figure}[ht]
\begin{center}
\includegraphics[height=3.3cm]{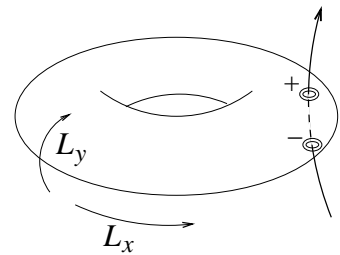}
\caption{A 2D superfluid with vortex excitations.  The vortices are
located at the points where the flux tube of circulation pierces the 
surface of the torus ($+$ and $-$ refer to a vortex and an antivortex).
}
\label{torus1}
\end{center}
\end{figure}

\section{Stiffness Parameters}
\label{PSsec}

\begin{figure}[ht]
\begin{center}
\includegraphics[height=1.9cm]{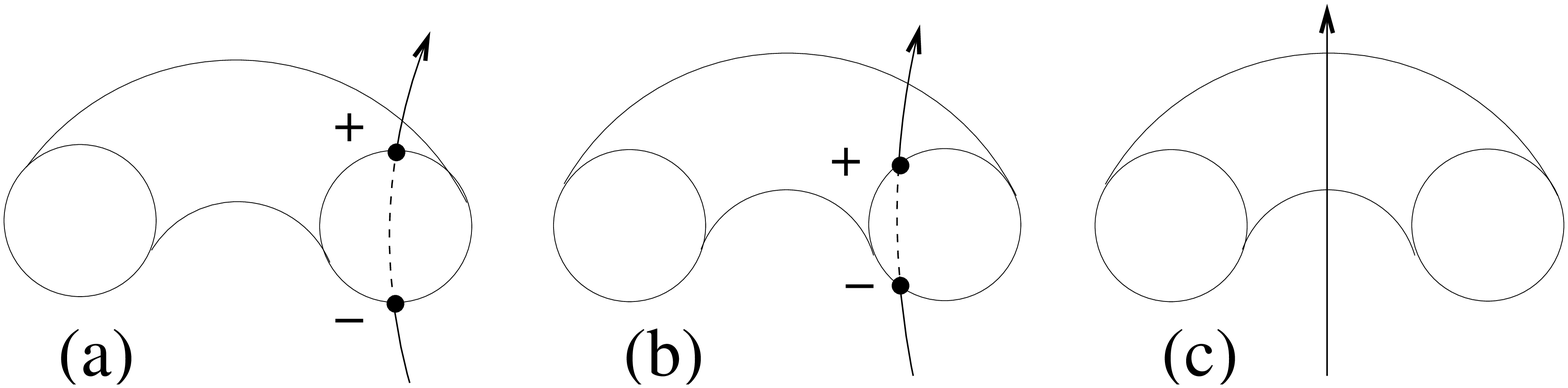}
\caption{Cross section of a torus, showing a flux line as it passes
through to the center of the superfluid.  In (a) the vortex-antivortex
pair is maximally separated, in (b) the pair moves closer together and in
(c) the pair has annihilated each other. The tube of quantized flux
remains.}
\label{torus3}
\end{center}
\end{figure}

Consider the $L_x\times L_y$ torus shown in
Fig.~1. Imagine that its surface is coated with a 2D
superfluid and that a tube of quantized circulation $h/m$
penetrates the torus as shown in Fig.~1. 
There is an antivortex in the superfluid layer at the point where this
flux tube passes into the torus, and a vortex at the point where it
leaves the surface.  These excitations
are the Kosterlitz-Thouless vortex-antivortex pairs.
On a finite torus one can also envision a situation in which the flux
tube moves through the cross-section
and into the center of the torus as illustrated in Fig.~\ref{torus3}.
In this case a single unit of quantized flux has entered the torus
(Fig.~\ref{torus3}(c)), and the excitation energy is 
\begin{equation}
\epsilon_x = \frac{\rho_s(T)}{2}
\ \left(\frac{2\pi}{L_x}\right)^2 L_x\, L_y = 2\pi^2
\rho_s(T)\ \frac{L_y}{L_x}.
\label{four}
\end{equation}  
Here, $\rho_s(T)$ is the usual Kosterlitz-Thouless
superfluid density which is renormalized by the
vortex-antivortex pairs.  Naturally, there are
excitations with energy $\epsilon_x \ell^2$
associated with flux tubes containing $\ell$ quanta.
Similarly, there are excitations associated with a
flux tube which threads around the inside of the torus
along the $L_x$ direction. The energies of these
excitations will vary as $L_x/L_y$.

At finite temperatures, the phase stiffness
parameters $\rho^W_x$ and $\rho^W_y$ are affected by the vortex
excitations. In the usual way, the 
stiffness is determined from the change in the free
energy associated with an infinitesimal flux $\phi$
threading the torus.  With $\phi$ in the direction of
the flux tube in Fig.~1, one obtains $\rho^W_x(T)$ from
\begin{equation}
\Delta F_x = \frac{1}{2}\ \rho^W_x(T)
\ \frac{L_y}{L_x}\ \theta^2 ,
\label{five}
\end{equation}
with $\theta = 2\pi \phi/(h/m)$. As discussed in
Ref.~\onlinecite{PSxx}, a calculation of $\Delta F$ gives 
\begin{equation}
\rho^W_x(T)  =  \rho_s(T)\ \left(1 -
\frac{4\pi^2\rho_s(T)}{T}\ \frac{L_y}{L_x}
\langle\ell^2\rangle_x\right), 
\label{six}
\end{equation}
where $\ell$ is the number of quanta in the 
flux tube, and
\begin{equation}
\langle\ell^2\rangle_x  =  \frac{\sum\limits_\ell
\ e^{-\left(\frac{\epsilon_x}{T}\right)\ell^2}
\ell^2}{\sum\limits_\ell\ e^{-
\left(\frac{\epsilon_x}{T}\right)\ell^2}}.
\label{seven}
\end{equation}
The stiffness in the y-direction is obtained
by replacing $L_y/L_x$ by $L_x/L_y$ in
Eqs.~(\ref{six}) and (\ref{seven}).

If one lets $L_x$ go towards infinity, keeping $L_y$
finite,
\begin{equation}
\langle\ell^2\rangle_x \cong
\frac{(L_x/L_y)\, T}{4\pi^2\rho_s(T)},
\label{eight}
\end{equation}
so that $\rho^W_x$ goes to zero as expected for a 1D
system at finite $T$. Alternatively, if the aspect
ratio is such that $L_x$ is large compared to $L_y$,
then
\begin{equation}
\rho^W_y(T) = \rho_s(T)\ \left(1 - 
\frac{8\pi^2 \rho_s(T)}{T}
\ \frac{L_x}{L_y}\ e^{- 2\pi^2 \frac{\rho_s(T)}{T}
\ \frac{L_x}{L_y}}\right),
\label{nine}
\end{equation}
and the difference between $\rho^W_y(T)$ and
$\rho_s(T)$ vanishes exponentially as $L_x/L_y$
increases.
Finally, in 3D the additional factor of $L_z$ which occurs in
Eq.~(\ref{six})
assures the convergence of $\rho^W_x$ to $\rho_s$ for all $R$
as the size of the system goes to infinity.

In the 2D superfluid, the jump in the stiffnesses at the 
Kosterlitz-Thouless temperature also clearly depends on the aspect ratio. At
$T_{KT}$, using the Nelson-Kosterlitz relation (\ref{tkt}), one has
\cite{PSxx}
\begin{equation}
\rho^W_x(T_{KT}) = \rho_s(T_{KT})\
\ \left( 1 - 8\pi\ \frac{L_y}{L_x}\ \langle \ell^2
\rangle_x\right),
\label{ten}
\end{equation}
with 
\begin{equation}
\langle\ell^2 \rangle_x  =  \frac{\sum \ell^2 e^{-4\pi \frac{L_y}{L_x} \ell^2} }
{\sum e^{-4\pi \frac{L_y}{L_x} \ell^2}}.
\label{twelve} 
\end{equation}
In a similar way, $\rho^W_y(T_{KT})$ and $\langle\ell^2 \rangle_y$ are obtained by replacing ${L_y}/{L_x}$ by ${L_x}/{L_y}$ in the above two
equations.
Hence, one can determine the ratios $\rho^W_x/ \rho_s$ and $\rho^W_y/ \rho_s$  
at $T_{KT}$ by evaluating the sums for $\langle\ell^2 \rangle_x$ and 
$\langle\ell^2 \rangle_y$.  Table~\ref{XYrat} shows the result of this 
calculation for various aspect ratios.

\begin{table}[ht]
\caption{Stiffness parameter ratios at $T_{KT}$, for different aspect
ratios $R = L_x / L_y$.  The entry * was evaluated
explicitly, \cite{Berndt}  while the remaining entries
were evaluated to arbitrary numerical accuracy 
using Waterloo's Maple V software.
}
\label{XYrat}
\begin{center}
\begin{tabular}{cp{5mm}cp{5mm}c}
$R$ && ${\rho^W_x}/{\rho_s}$ &&
${\rho^W_y}/{\rho_s}$ \\ \colrule
$1$ &&  $0.9998247$ &&  $0.9998247$ \\
$2$ &&  $0.9532407$ &&  $1 - 1.222613\times 10^{-9}$ \\
$3$ &&  $0.7533903$ &&  $1 - 6.395505\times 10^{-15}$ \\
$4$ &&  $1/2^*$     &&  $1 - 2.973776\times 10^{-20}$\\
$5$ &&  $0.2977676$ &&  $1 - 1.296322\times 10^{-25}$ \\
$6$ &&  $0.1663429$ &&  $1 - 5.424861\times 10^{-31}$ \\
$7$ &&  $0.0893401$ &&  $1 - 2.207141\times 10^{-36}$ \\
$8$ &&  $0.0467593$ &&  $1 - 8.796634\times 10^{-42}$
\end{tabular}
\end{center}
\end{table}

\section{Monte Carlo Calculations}
\label{MCsec}

In order to study this aspect ratio dependence of $\rho^W_x$ and
$\rho^W_y$ we employ
a model of a two dimensional superfluid using a 
hard-core boson Hamiltonian at half filling, which is equivalent to the quantum
$S = {1}/{2}$ XY model defined by 
\begin{equation}
H = -J \sum_{\langle i,j\rangle} \left[{S_i^x S_j^x + S_i^y S_j^y
}\right].
\label{XYhamil}
\end{equation}
Here $S_i^x$ and $S_i^y$ are the $x$ and $y$ components of a spin
${1}/{2}$ operator at site $i$, and the sum is carried out over
all nearest-neighbor spin pairs $\langle i,j\rangle$.  
This model is known to have a Kosterlitz-Thouless type transition from 
previous Monte Carlo simulations carried out on square lattices.
\cite{AWS1, Harada}
We carry out simulations using the stochastic series 
expansion (SSE) quantum Monte Carlo method, \cite{AWS3} 
that has previously been
applied to this and other spin and boson models.  The basis of the SSE
method is importance sampling of the power series expansion of the partition 
function:
\begin{equation}
Z = Tr\{ e^{-\beta H} \} 
= \sum_{\gamma} \sum_{n=0}^{\infty} \frac{(-\beta)^n}{n!} 
\langle{\gamma| H^n | \gamma} \rangle ,
\label{SSEpart}
\end{equation}
where $\beta = 1/T$ is the inverse temperature and
the trace has been written as a sum over diagonal matrix elements
in a basis $\{ |\gamma\rangle\}$. In our case this is 
the standard basis of $z$ spin quantization.  In the simulations we
employ an efficient cluster-type method for sampling the terms
of the expansion, called the {\it directed-loop algorithm}, \cite{AWS3}
which in the present case of zero field is similar to loop algorithms 
previously employed on the spin ${1}/{2}$ XY model.  \cite{Harada}
Note that all contributing expansion-orders $n$ in (\ref{SSEpart}) 
are sampled, and all results are exact within statistical errors.

A direct estimator for the superfluid (spin) stiffnesses is identified 
in the SSE quantum Monte Carlo in a way analogous to world-line quantum
Monte Carlo methods. \cite{stiffness, Harada}
Starting from the definition Eq.~(\ref{five}) and taking the second 
derivative of the free energy (per spin) with respect to a twist
$\theta$ in the boundary condition 
will lead to the expressions given in Eq.~(2) and (\ref{three}). 
The winding number in these equations is now a measure of the
net spin currents flowing around the periodic system, 
$W_{\alpha} = (N_{\alpha}^R - N_{\alpha}^L) / L_{\alpha}$,
($\alpha=x,y$)
where $N_{\alpha}^{R,L}$ is the number of operators transporting spin
to the ``right'' or ``left'' along the $\alpha$ direction in the SSE
configuration.

\subsection{Spin Stiffness in 2D and 3D}
\label{23Dsec}

\begin{figure}[ht]
\begin{center}
\includegraphics[height=9cm]{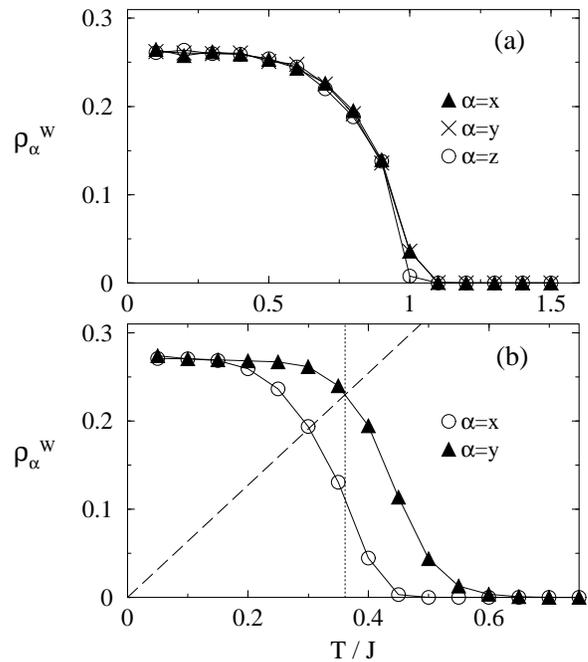}
\caption{ The spin stiffness order parameter Eqs.~(2),(\ref{three})
calculated for a (a) 16$\times$16$\times$64 system and a
(b) 64$\times$16 system.  The 
long lattice direction is $z$ and $x$, respectively.  The slight
deviation of $\rho^W_z(T)$ around $T/J \approx 1.0$ in (a) disappears in the
limit of large system size.  The thick dashed line in (b) represents the 
equation $\rho^W_{\alpha} = {2T}/{\pi J}$,  
and the thin vertical dotted line marks the corresponding estimate of 
$T_{KT}/J$. 
}
\label{fss}
\end{center}
\end{figure}

In order to test the predictions of Prokof'ev and
Svistunov, \cite{PSxx}
we carried out a series of simulations using the Hamiltonian,
Eq.~(\ref{XYhamil}), for 
systems with various aspect ratios in two and three dimensions.  
In general our calculations confirm the predictions of Ref.~\onlinecite{PSxx},
namely that in the 3D system, the difference between 
the components of $\rho^W_{\alpha}(T)$ ($\alpha=x,y,z$) vanishes
exponentially, while in 2D, $\rho^W_x(T)$ can differ
significantly from $\rho^W_y(T)$ depending on the value of the aspect
ratio $R$.
Specific examples are illustrated in Fig.~\ref{fss}.  As we see in
Fig.~\ref{fss}(a), in the 3D case, 
for $L_x/L_y = 1$ and $L_z/L_y = 4$ 
the three different spin stiffness parameters converge to one value.
In contrast, for the 2D case (Fig.~\ref{fss}(b)), 
with $R=L_x/L_y = 4$ the spin stiffness in the long
direction is significantly less that the spin stiffness in
the short direction for a large range of $T$.
As we will see below, in 2D $\rho^W_x(T)$ does not converge to 
$\rho^W_y(T)$ in this temperature range even in the limit of large 
system sizes (over $6.5\times 10^4$ spins).

A rough estimate of $T_{KT}$ and the ratio 
$\rho^W_x(T_{KT}) / \rho^W_y(T_{KT})$ can be obtained from
Fig.~\ref{fss}(b),
by drawing a straight line $\rho^W_{\alpha} = {2T}/{\pi J}$.  
The point where this line intersects $\rho^W_y \approx \rho_s$
is a finite-size measurement of $T_{KT}/J$, and the ratio $\rho^W_x / \rho^W_y$ 
at this $T$ can be compared to the value 1/2 from Table~\ref{XYrat}
(which we expect to be exact in the limit of infinite system size).
It is clear from this simple demonstration that if one wishes to 
study the spin stiffness dependence on $R$ in detail, an accurate 
independent estimate of $T_{KT}$ will facilitate quantitative comparisons of 
$\rho^W_x / \rho^W_y$ with the entries in Table~\ref{XYrat}.
The next section of this paper is therefore dedicated to the
measurement of $T_{KT}$.

\subsection{Weber-Minnhagen\cite{WMxx} determination of $T_{KT}$}
\label{WMsubsec}

We now give details on the independent estimate for $T_{KT}$ 
%using the Weber-Minnhagen\cite{WMxx} scaling for the system size 
%dependence of the spin stiffnesses, 
which will be 
used as a benchmark for $T_{KT}(L \rightarrow \infty)$ in the discussion
to follow.  Our approach follows that of
Harada and Kawashima, \cite{Harada} who carried out simulations of the quantum
$S = {1}/{2}$ XY model on nine square lattices of size $L=8$ to 128.
Their analysis found a $T_{KT} = 0.3427(2)J$, based on a Weber-Minnhagen
\cite{WMxx} scaling fit for the system size dependence of the spin
stiffness.  Specifically, this scaling form states that as $T \rightarrow T_{KT}^-$,
\begin{equation}
\rho_s(L) = \rho_s(\infty) \left[{1+ \frac{1}{2}
\frac{1}{\ln (L) + C)} }\right] ,
\label{WMfit}
\end{equation}
where $\rho_s(L)$ now represents the spin stiffness at a finite
lattice with a linear size of $L$ spins,
$\rho_s(\infty)$ is given by Eq.~(\ref{tkt}),
and $C$ is some unknown constant.
One can also reformulate Eq.~(\ref{WMfit}) in the linear form,
\begin{equation}
Y(L) = \pi\ln (L) + C,
\label{RGMfit}
\end{equation}
where
\begin{equation}
Y(L) = \left[{\frac{\rho^W_y(L)}{F(R)} \frac{1}{T}-\frac{2}{\pi}}\right]^{-1}, 
\label{LINfit}
\end{equation}
and $C$ has been re-scaled to absorb some
simple constants.  The factor $F(R)$ has been included in order to
correct $\rho_y^W$ according to $\rho_s= \rho ^W_y /F(R)$ for square systems,
where $F(1) = 0.9998247$ from Table~\ref{XYrat}.

\begin{figure}[ht]
\begin{center}
\includegraphics[height=5.5cm]{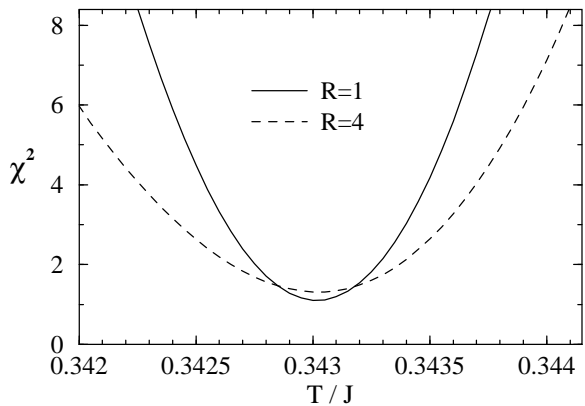}
\caption{
$\chi ^2$ values (per degree of freedom) for $R=1$, lattice sequence
$L = 16,24,32,40,48,56,64,72,96,128$ and
$R=4$, $L_y = 16,20, 24,32,40,48,56,64$.
}
\label{Chi2Slope}
\end{center}
\end{figure}

In order to obtain results on a dense temperature grid,
we did an extensive series of  {\it simulated tempering} \cite{MARxx}
simulations within a $T$ range centered about the approximate
$T_{KT} \approx 0.343J$. \cite{Harada}
Fits to Eq.~(\ref{RGMfit}), adjusting $C$ only,
should have a minimum in $\chi ^2$ at $T_{KT}/J$.
This is illustrated in Fig.~\ref{Chi2Slope}.   In general, we find that
the qualities of the $\chi ^2$ curves can differ significantly depending
on the ``lattice sequence'' included in the fit.  An acceptable lattice
sequence should be one that produces a $\chi ^2$ minimum of
approximately unity.  We therefore systematically eliminate the smallest $L_y$
values from the lattice sequence until $\chi ^2_{min} \approx 1$.
With our tempering data, this occurs when the largest excluded data
point is $L_y=8$, however to be cautious we also excluded $L_y=12$.
Qualitatively, the shift in $\chi ^2_{min}$ is very small
($\Delta T/J < 0.0001$) when $L_y>16$ points are excluded from the data
set, which in part is due to the trivial statistical systematic shift resulting from 
the elimination of data points from the fit.
Finally, the $\chi^2$ curves have little quantitative dependence on the
maximum $L_y$ included, as long as this $L_y \geq 64$.

\begin{figure}[ht]
\begin{center}
\includegraphics[height=7.3cm]{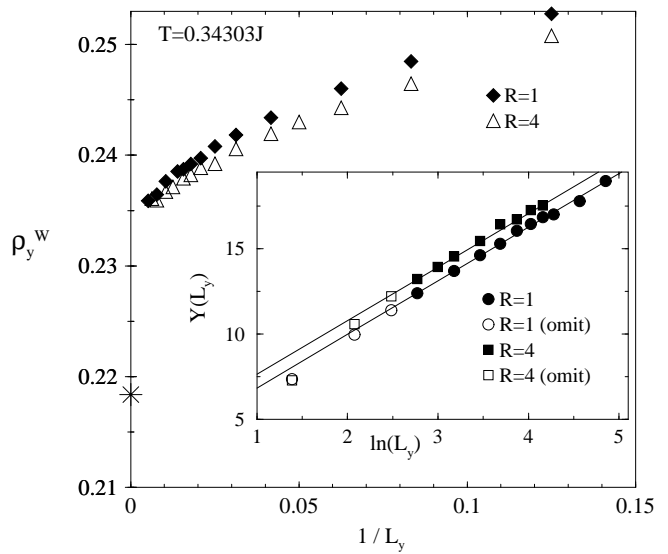}
\caption{
Simulated tempering data for $\rho^W_y$, where each data point represents 
$2 \times 10^7$ Monte Carlo production steps.
Statistical errors are at most of the order of the symbol sizes.
The star on the vertical axis is $\rho_y^W(L_y \rightarrow \infty)$
determined from $T_{KT} = 0.343J$ using Eq.~(\ref{tkt}).
Inset: $\rho^W_y$ data scaled to the form Eq.~(\ref{RGMfit}). 
Data included in the $\chi ^2$ fit is plotted with solid
symbols, and data for smaller system sizes that was omitted from the
$\chi ^2$ fit is plotted with open symbols.  
The solid lines have constant slopes of $\pi$.
}
\label{TmprRHO}
\end{center}
\end{figure}

Using the scaling form Eq.~(\ref{RGMfit}) with the data illustrated in
Fig.~\ref{TmprRHO}, we obtain three estimates for $T_{KT}(\infty)$:
for (i) $R=4$, using $F(R)=1$, (ii) $R=1$, with the approximation $F(R)=1$,
and (iii) $R=1$, with the correct factor $F(R)=0.9998247$ 
from Table~\ref{XYrat}. The corresponding results for the
transition temperatures are
$T^{(i)}_{KT}/J=0.34302 \pm 1.4\times 10^{-4}$,
$T^{(ii)}_{KT}/J=0.34302 \pm 0.9\times 10^{-4}$ and
$T^{(iii)}_{KT}/J=0.34304 \pm 1.0\times 10^{-4}$, respectively,
where quoted errors are one standard deviation.  From this, we see that
$T^{(i)}_{KT}= T^{(ii)}_{KT}=T^{(iii)}_{KT}$
to within our statistical error.  To obtain a more accurate estimate for
the transition temperature, we may also perform a weighted average of our
values for $T^{i}_{KT}$ and $T^{iii}_{KT}$, to get
$T_{KT}/J=0.34303(8)$.
This value agrees to the previous\cite{Harada} most accurate 
$T_{KT} = 0.3427(2)J$ to within two standard deviations.

The Monte Carlo data and the $\chi^2$ fit at our estimated $T_{KT}$ are shown 
in Fig.~\ref{TmprRHO}.  As evident there, the unscaled $\rho_y^W$ data     
at $T/J = 0.34303$ do not clearly approach the infinite size limit
determined with Eq.~(\ref{tkt}).  However, when properly scaled according
to Eq.~(\ref{LINfit}), $Y(L)$ does become linear according to the expected form
Eq.~(\ref{RGMfit}) (Fig.~\ref{TmprRHO} inset).

\subsection{The Spin Stiffness Ratio $\rho_x^W / \rho_y^W$ at $T_{KT}$}
\label{rhoRATsec}

In this section we study the aspect ratio dependence of the spin stiffness
in more general terms, and outline a procedure for calculating $T_{KT}$
using $\rho_x^W/\rho_y^W$ and the results of Table~\ref{XYrat}. Our aim
here is not to obtain a more precise value for $T_{KT}$, but to test how
closely the predicted values for the ratios can be observed in practice.

To begin,
we collected detailed data for $\rho^W_x(T)$ and
$\rho^W_y(T)$ for 20 to 24 values of $T/J$ between 0.325 and 0.375.
Simulations produced between $1 \times 10^6$ and $5 \times 10^6$
Monte Carlo steps at each $T$ (depending on system size and statistical 
error bars), except for the very largest systems which had an order of 
magnitude less Monte Carlo steps.  System aspect ratios included were $R=$1,2
and 4.  Continuous data over $T$ was obtained by fitting third-order polynomial
curves.

\begin{figure}[ht]
\begin{center}
\includegraphics[height=8.3cm]{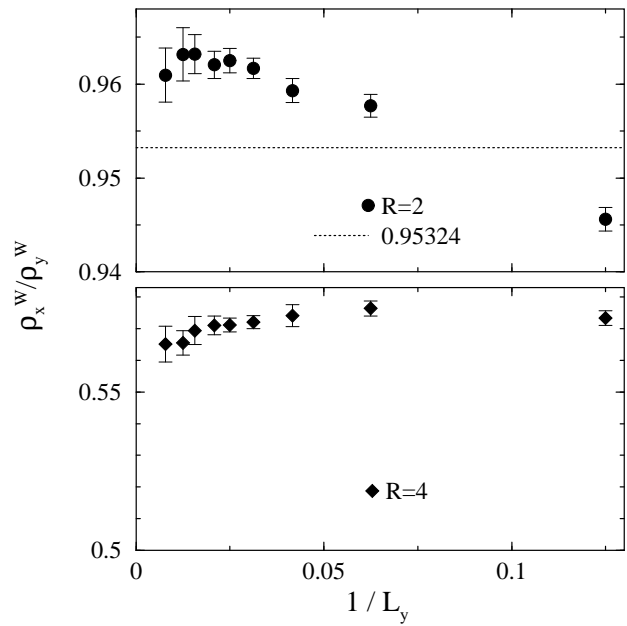}
\caption{
The spin stiffness ratios for $R=2$ and $R=4$ at $T/J=0.34303$.
At $T=T_{KT}$ and $L_x,L_y \rightarrow \infty$, the values are expected
to approach $\rho_x^W/\rho_y^W=0.95324$ and $1/2$, respectively.
}
\label{T0p343R2n4rats}
\end{center}
\end{figure}

\begin{figure}[ht]
\begin{center}
\includegraphics[height=6.2cm]{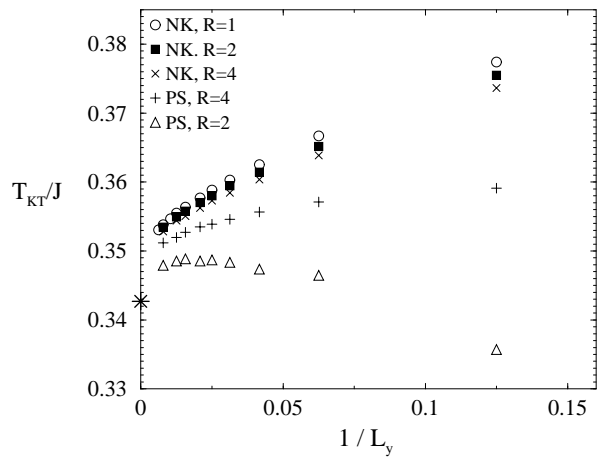}
\caption{
The Kosterlitz-Thouless transition temperature calculated with the
Nelson-Kosterlitz formula (NK, Eq.~(\ref{tkt})), and
with the spin stiffness aspect ratios (PS, Table~\ref{XYrat}),
 as a function of inverse linear system size.
The infinite size limit is $T_{KT} \approx 0.343 J$ 
(see Section~\ref{WMsubsec} of text) 
and is plotted as a star on the vertical axis.
}
\label{SCALEcomb}
\end{center}
\end{figure}

Fig.~\ref{T0p343R2n4rats} shows the spin stiffness ratios 
$\rho_x^W/\rho_y^W$ calculated with this data, 
at $T_{KT}/J=0.34303$, as was determined independently in 
Section~\ref{WMsubsec} above. 
As evident in the figure, this type of simple analysis is not
very satisfying, as the spin stiffness ratios do not obviously
approach the values expected from Table~\ref{XYrat}. 
Clearly there are significant finite-size effects at play,
as seen previously in the Weber-Minnhagen scaling analysis of
Section~\ref{WMsubsec} (Fig.~\ref{TmprRHO}).
One way to quantitatively account for the scaling effects
in our $\rho_x^W/\rho_y^W$ will now be outlined.
To do so, we define a new measurement technique which can be used
to estimate the Kosterlitz-Thouless transition temperature: namely, 
$T_{KT}$ is defined as the temperature at which the ratio $\rho^W_x/\rho_s$ 
(in practice $\rho^W_x/\rho^W_y$), passes through the corresponding 
value in column 2 of Table~\ref{XYrat} (for $R=$2 and 4 only).
Additionally, for the purposes of comparison, $T_{KT}$ can be obtained by 
calculating the crossing of Eq.~(\ref{tkt}) with the $\rho^W_y(T)$ 
data for each system size $L_x \times L_y$.  
The results of these two procedures are summarized in Fig.~\ref{SCALEcomb}.

Several interesting observations can be made from this figure.  First,
if one were only to consider smaller system sizes ($L_y < 64$), the
data would appear to consistently approach an infinite size limit
of $T_{KT}/J \approx 0.35$. 
However, at $L_y \approx 64$ we observe that all of the data sets in 
Fig.~\ref{SCALEcomb} undergo subtle changes in slope that could
indeed suggest a significantly lower value for $T_{KT}$ in the infinite size 
limit. Note in particular the change in the sign of the slope for 
the PS, $R=2$ line at $L_y \approx 64$. Without any {\it a priori} 
knowledge of the form of the scaling laws for the curves in 
Fig.~\ref{SCALEcomb}, it would be difficult to conclude whether 
they approach the 
Kosterlitz-Thouless temperature of $T_{KT} \approx 0.343J$
calculated in Section~\ref{WMsubsec}, and in Ref.~\onlinecite{Harada},
illustrated on the vertical axis of Fig.~\ref{SCALEcomb}.

\begin{figure}[ht]
\begin{center}
\includegraphics[height=9cm]{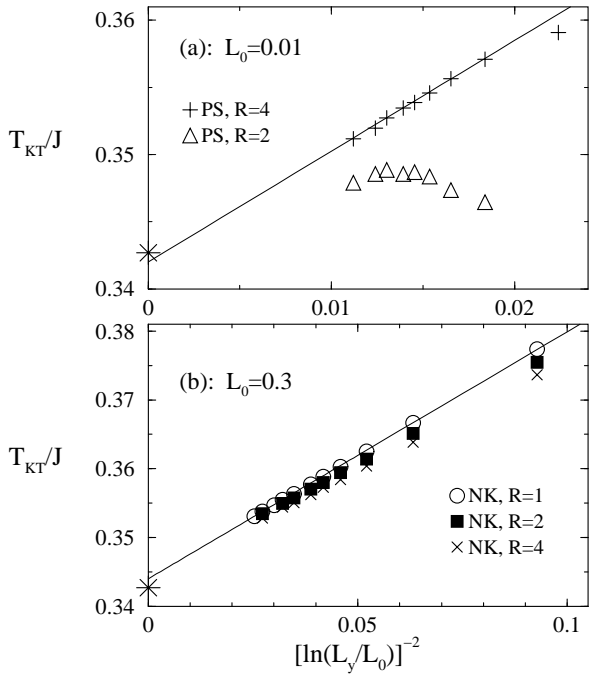}
\caption{
The Kosterlitz-Thouless transition temperatures scaled according to
Eq.~(\ref{KRGfit}).  In (a), the $T_{KT}$ data for $R=4$ becomes approximately
linear (excluding the smallest system size, $L_y=8$) for $L_0 \approx
0.01$.  In (b), all $T_{KT}$ curves become approximately linear for $L_0$ in a
narrow range around 0.3.  Straight lines are included as guides for the eye.  
The infinite size limit is $T_{KT} = 0.343 J$ (see Section~\ref{WMsubsec} of 
text) and is plotted as a star on the vertical axis.
}
\label{APYscaleFig}
\end{center}
\end{figure}

We now consider a scaling law for $T_{KT}$ that arises from 
the Kosterlitz RG scaling equations.\cite{KosterlitzRG}
As discussed in Ref.~\onlinecite{KosterlitzRG}, the functional form of the
correlation length suggests that, as $T \rightarrow T_{KT}$, $\xi$
diverges according to
\begin{equation}
\xi \sim \textrm{exp}(bt^{-1/2}),
\end{equation}
where $t=(T-T_{KT})/T_{KT}>0$ and $b$ is a constant.   If we identify   
$\xi=L/L_0$ ($L_0$ some microscopic length), and $T=T_{KT}(L)$, then 
we can write the system size-dependent Kosterlitz-Thouless transition
temperature in terms of $L/L_0$ as
\begin{equation}
T_{KT}(L) = T_{KT}(\infty)\left[{ 1 + \frac{b^2} {\ln^2(L/L_0)}
}\right].
\label{KRGfit}
\end{equation}
Fig.~\ref{APYscaleFig} illustrates the $T_{KT}(L)$ data of
Fig.~\ref{SCALEcomb}, re-scaled to the form of Eq.~(\ref{KRGfit}), for some
value of $L_0$ which approximately linearizes the data.  
As this figure shows,    
by choosing appropriate values for $L_0$ it is possible to obtain 
linear fits to  
Eq.~(\ref{KRGfit}) for all data sets save one.  The exception is the 
spin stiffness aspect ratio crossing for $R=2$ (PS, $R=2$ in 
Figs.~\ref{SCALEcomb}
and \ref{APYscaleFig}).   In this case, the change in slope of the data set
at $L_y \approx 64$ precludes any direct fitting to the scaled form
Eq.~(\ref{KRGfit}).  However, the data for $L_y > 64$ show 
rough evidence of this scaling trend, and it is therefore very likely that 
additional data for $L_y > 128$ would approach this $T_{KT}(\infty)$ 
consistently.

As Fig.~\ref{APYscaleFig} shows,
the approximate intercept of these scaled data sets are in good agreement
with the value of $T_{KT}(\infty)$ calculated in Section~\ref{WMsubsec}
using the Weber-Minnhagen\cite{WMxx} scaling form
for the spin stiffness.  In particular, the 
consistency of the intercept for the data set PS, $R=4$, adds          
confidence to the assertion that $T_{KT}(L)$ calculated using the
spin stiffness aspect ratios is a good estimator for the Kosterlitz-Thouless 
transition temperature, consistent with $T_{KT}$ calculated using other 
methods. Thus the entries in Table~\ref{XYrat} indeed appear to be
accurate. Clearly, however, our value of
$T_{KT}/J=0.34303(8)$ obtained from the Weber-Minnhagen scaling is the
most accurate because it is based on a one-parameter fit,
Eq.~(\ref{RGMfit}), as opposed to the two-parameter fit,
Eq.~(\ref{KRGfit}).

\section{Conclusions}

In conclusion, we have confirmed the prediction of Prokof'ev and 
Svistunov \cite{PSxx} that in 2D the squared winding-number estimates
of the finite-temperature 
spin stiffnesses $\rho^W_x$ and $\rho^W_y$
differ from each other for $R \neq 1$, while in 3D systems 
$\rho^W_x = \rho^W_y = \rho^W_z$ for any $R$.
In 2D we also find that $\rho^W_y$ approaches the Nelson-Kosterlitz
superfluid density $\rho_s$ exponentially for
$R>1$ as $L_x,L_y \rightarrow \infty$, whereas $\rho^W_x$ approaches
zero as $R \to \infty$ exactly as predicted.

As illustrated in Fig.~\ref{APYscaleFig}, 
the $T_{KT}$ predicted from the ratios
$\rho^W_x / \rho^W_y$ approach the $T_{KT}$ predicted using other methods,
substantiating the validity of the results listed in Table~\ref{XYrat}. 
Neglecting the small differences between $\rho_s$ and $\rho^W_{x,y}$ on 
2D $L\times L$ ($R=1$) lattices is insignificant 
compared to presently typical statistical
uncertainties. However, with any further increase in accuracy relative
to that achieved in the data presented here, the effect has to be
taken into account in order to avoid a systematic error.

\acknowledgments{
The authors would like to thank L.~Balents and A.~P.~Young for insightful 
discussions.  
Supercomputer time was provided by NCSA under grant number DMR020029N,
and the UCSB Materials Research Laboratory.
Financial support was provided by National Science Foundation Grant No.
NSFDMR98-17242 (DJS) and the Academy of Finland, project No.~26175 (AWS).
AWS would also like to thank the UCSB physics department for hospitality
and support during a visit}.

\end{document}